\input{aipcheck}

\documentclass[
    ,final            % use final for the camera ready runs
  ]
  {aipproc}

\layoutstyle{8x11single}

\usepackage{amsmath,amssymb}

\newcommand{\ie}{\mbox{\it i.e.}}
\newcommand{\eg}{\mbox{\it e.g.}}

\newcommand{\ben}{\begin{eqnarray}}
\newcommand{\een}{\end{eqnarray}}
\newcommand{\nn}{\nonumber}
\newcommand{\wt}{\widetilde}
\newcommand{\citeeq}[1]{\mbox{Eq. (\ref{#1})}}
\newcommand{\mchi}{\mbox{$m_{\chi}$}}
\newcommand{\sigv}{\mbox{$\langle \sigma v\rangle$}}
\newcommand{\prob}{\mbox{${\cal P}$}}
\newcommand{\msun}{\mbox{$M_\odot$}}
\newcommand{\cl}[1]{\mbox{${#1}_{\rm cl}$}}
\begin{document}

\title{Cosmic ray positron excess:\\ is the dark matter solution a good bet?}

\classification{95.35.+d;12.60.-i;95.30.Cq;96.50.S-}
\keywords      {Dark matter; Galactic cosmic rays}

\author{Julien Lavalle}{
  address={Universit\`a di Torino \& INFN, Dipartimento di Fisica Teorica, 
  Via Giuria 1, 10125 Torino --- Italia},
  email={lavalle@to.infn.it}
}

\begin{abstract}
The recent observation by the PAMELA satellite of a rising positron fraction 
up to $\sim$ 100 GeV has triggered a considerable amount of putative 
interpretations in terms of dark matter (DM) annihilation or decay. Here, we 
make a critical reassessment of such a possibility, recalling the elementary 
conditions with respect to the standard astrophysical background that would 
make it likely, showing that they are not fulfilled. Likewise, we argue that, 
as now well accepted, DM would need somewhat contrived properties to 
contribute significantly to the observed positron signal, even when including 
\eg~clumpiness effects. This means that most of natural DM candidates 
arising in particle physics beyond the standard model are not expected to be
observed in the cosmic antimatter spectrum, unfortunately. However, this 
does not prevent them from remaining excellent DM candidates, this only 
points towards the crucial need of developing much more complex detection 
strategies (multimessenger, multiwavelength, multiscale searches).
\end{abstract}

\maketitle

\section{Introduction}
\label{sec:intro}
Since its discovery in the early 1930's by \citet{1933AcHPh...6..110Z}, 
the DM issue has remained unsolved. There are basically two different 
theoretical ways to address this issue, one considering a new additional 
component of exotic matter in the form of weakly interacting massive particles 
(WIMPs), the other involving modifications of general relativity (sometimes 
even both). Both have relevant motivations, the former from particle physics 
beyond the standard model~\citep[see \eg][for reviews]{1996PhR...267..195J,2007arXiv0704.2276M} and structure formation~\citep[see a more detailed discussion in \eg][]{2009arXiv0910.5142P}, the latter from more empirical attempts at the 
galaxy scale~\citep{1983ApJ...270..365M} or more recently in the context of 
extra-dimensional theories~\citep[\eg][]{2000PhLB..485..208D}. One of the 
appealing flavor of the former hypothesis, that we will consider in the 
following, is the possibility to test it with a broad variety of existing 
or coming experimental devices. Among interesting 
\emph{astroparticle} signatures, gamma rays and antimatter CRs have long been 
considered as promising DM tracers~\citep{1978ApJ...223.1015G,1984PhRvL..53..624S}, but it is only recently that precision data have become available to 
look for non-standard features~\citep{2006RPPh...69.2475C,Salati:2007zz}.

Although the rise in the local cosmic positron fraction at the GeV energy scale 
has been observed for a long time \citep[\eg][]{1969ApJ...158..771F,1997ApJ...482L.191B,2000PhLB..484...10A}, the statistics recently released by the 
PAMELA collaboration~\citep{2009Natur.458..607A} is unprecedented and covers 
a much larger energy range, up to 100 GeV. The secondary origin of these 
positrons seems unlikely~\citep{1998ApJ...493..694M,2009A&A...501..821D}, even 
if theoretical uncertainties are still large. The main questions are therefore 
(i) whether or not standard astrophysical sources may supply for such a signal 
and (ii) whether or not DM annihilation or decay is expected to be also 
observed in this channel. It is noteworthy that this was already discussed 
by~\citet{1989ApJ...342..807B} twenty years ago, where the author pointed out 
that a pulsar origin was the best explanation to a rising positron fraction. It 
is not less interesting and sociologically striking to take a census of the 
articles addressing point (i) versus those focused on point (ii). Anyway, in 
this proceeding, we aim at discussing this issue concentrating on the local 
cosmic positron signal only, forgetting about other counterparts. We will first 
review the astrophysical backgrounds of secondary and primary origins; then, we 
will check whether DM could naturally yield prominent imprints in the cosmic 
positron spectrum, before concluding.

\section{Astrophysical backgrounds}
\label{sec:astro}

\subsection{Bases of CR propagation}
The global understanding of Galactic CRs at the GeV-TeV scale is rather 
well established. CRs are accelerated by shock waves at the vicinity of violent 
events like supernova (SN) explosions, and further diffuse erratically in the 
interstellar medium (ISM) by bouncing on moving magnetic turbulences. This 
diffusive motion is accompanied by other processes whose respective impacts 
depend on the cosmic ray species: convection that drives CRs away from the 
Galactic plane (negligible above a few GeV), energy losses (affecting mostly 
leptons), diffusive reacceleration (negligible above a few GeV), 
spallation reactions with the ISM gas (for nuclei only). The general formalism
of CR transport was designed a long time ago in the seminal 
book of~\citet{1964ocr..book.....G}, and refined many times since 
then~\citep[see \eg][]{berezinsky_book_90,2007ARNPS..57..285S}. The master
equation that describes the CR transport in phase-space looks like a classical 
current conservation equation:
\ben
\widehat{\cal D}_\mu \,\widehat{\cal J}^\mu + 
\widehat{\cal D}_E \,\widehat{\cal J}^E = \widehat{\cal Q}\;.
\label{eq:transport}
\een
Given a CR differential number density ${\cal N}=dn/dE$, the spacetime-like 
current is reminiscent from the Fick law and the heat equation, 
$\widehat{\cal J}^\mu = ({\cal N}, \left\{\vec{\nabla} K_d(E) - 
\vec{V}_c\right\}{\cal N})$, for which the associated transport operator reads 
$\widehat{\cal D}_\mu = (\partial_t+\Gamma_{\rm spal}+\Gamma_{\rm dec}, -
\vec{\nabla})$. The energy component is merely $\widehat{\cal J}^E = {\cal N}$, 
on which acts $\widehat{\cal D}_E = \partial_E (dE/dt +  K_E\partial_E)$. 
Appearing above, $\widehat{\cal Q}$ is the source term, $K_d$ the spatial 
diffusion coefficient, $K_E$ the reacceleration coefficient, 
$\Gamma_{\rm spal/dec}$ the spallation/decay rate, $\vec{V}_c$ the convection 
velocity, and $dE/dt$ the energy loss term. Each of these ingredients is by 
itself subject of intense researches, so that many simplifying assumptions are 
usually made in phenomenological analyses. In general, one assumes that spatial 
diffusion proceeds isotropically and that the diffusion coefficient is 
homogeneous in the diffusion zone, only scaling with the CR rigidity 
$\propto {\cal R}^\delta$. Apart from the energy losses and the spallation or 
decay rate which can be predicted independently, the propagation parameters are 
usually constrained with measurements of CR nuclei, more precisely with 
secondary to primary ratios like B/C
\citep[see \eg][]{1998ApJ...509..212S,2001ApJ...555..585M}. 
Important features of such a modeling are the spatial extent of the diffusion 
zone (usually taken as a cylindrical slab) and the diffusion coefficient, and 
we stress that the related uncertainties are still rather 
large~\citep{2001ApJ...555..585M}. In the 
following, we will adopt a thick cylindrical diffusion zone of radius $R=20$ 
kpc and half-width $L=4$ kpc, unless specified otherwise.

In some cases, analytical solutions to the diffusion equation can be found in 
terms of Green functions $\cal{G}$, which obey 
$\widehat{\cal D}_{\mu,E}\, {\cal G}^{\mu,E}(\widehat{x}\leftarrow 
\widehat{x}_s) = \delta^3(\vec{x}_s-\vec{x})\delta(t_s-t)\delta(E_s-E)$. For 
instance, assuming both steady state, which is relevant for a constant CR 
injection rate, and an infinite 3D diffusion space, thereby neglecting the 
spatial boundary conditions, the propagators for protons 
(or antiprotons) and electrons (or positrons) are simply given by:
\ben
{\cal G}_p(\vec{x}\leftarrow \vec{x}_s) = 
\frac{1}{4\pi K(E)||\vec{x}-\vec{x}_s ||}\;,\;\;\;
{\cal G}_e(E,\vec{x}\leftarrow E_s,\vec{x}_s) = 
\frac{ 1 }{b(E)\left\{ \pi \lambda^2 \right\}^{3/2}}\,
\exp\left\{-\frac{|\vec{x}-\vec{x}_s |^2}{\lambda^2}\right\}\;.
\label{eq:gp_ge}
\een
Only spatial diffusion has been considered for protons (no energy losses, 
convection, nor spallation --- fair approximation above a few GeV). For 
electrons, the mixed impact of (local) energy losses $dE/dt = -b(E)$ and of 
diffusion is encoded in the propagation scale 
\ben
\lambda(E,E_s) = 
\left\{4\int_E^{E_s} dE' \,K_d(E')/b(E')\right\}^{1/2}\;, 
\label{eq:lambda}
\een
and other processes are neglected. This illustrates that while proton and 
electron transport is described by the same equation, these species can 
actually diffuse quite differently because of the relative differences in the 
processes they experience. For electrons in 
the GeV-TeV range, energy losses are dominated by inverse Compton (IC) 
scattering on the interstellar radiation field (ISRF) including the cosmic 
microwave background (CMB), and by synchrotron losses on the Galactic magnetic 
field. In the local environment, the typical energy loss 
timescale at $E_0=1$ GeV is $\tau \simeq 315$ Myr. With a typical diffusion 
coefficient of $K_0\simeq 0.01\,{\rm kpc^2/Myr}$, one finds $\lambda\sim 3.5$ 
kpc, which justifies {\it a posteriori} the use of the local ISRF and magnetic 
field to compute the energy losses~\citep{2009A&A...501..821D}. In the Thomson 
approximation, $b(E) = (E_0/\tau) \left\{ \epsilon\equiv E/E_0\right\}^2$ 
\citep{1970RvMP...42..237B}, which implies that the propagation scale strongly 
decreases with energy.

The previous Green functions reveal an important difference between stable 
nuclei and electrons: the former have a long range propagation
scale (above a few GeV) only limited by the finite spatial extent of the 
diffusion zone, while the latter have a short range propagation scale limited 
by energy losses. Therefore, spatial fluctuations of the CR injection rate will 
be less important for stable nuclei (except below a few GeV, when spallation 
and convection become important) than for electrons. This implies a more local 
origin of high energy CR electrons and positrons, and means that time 
fluctuations in their local injection rate induces strong local effects.

\subsection{Astrophysical positrons}
Positrons of astrophysical origin can be secondaries or primaries. Secondaries
are produced from spallation reactions of CR nuclei (mostly protons and 
$\alpha$) with the ISM gas (H and He). Primaries can be directly produced in 
the intense magnetic field hosted by sources like 
pulsars~\citep{1969ApJ...157.1395O} and further accelerated in the surrounding 
shocked medium, or could also be secondaries created from spallation processes 
within acceleration sites like supernova remnants (SNRs)
\citep{2009PhRvL.103e1104B}. Thus, the secondary positron source term depends 
on the spatial distribution of CR nuclei and of the ISM gas:
\ben
{\cal Q}_s (E,\vec{x}) = 4\pi\sum_{i,j} 
\int dE' \,\phi_i(E',\vec{x}) \,\frac{d\sigma_{ij}(E',E)}{dE}\, 
n_j(\vec{x})\;,
\een
where $i$ flags the CR species of flux $\phi$ and $j$ the ISM gas species
of density $n$, the latter being concentrated within the thin Galactic disk. 
$d\sigma_{ij}(E',E)$ is the inclusive cross section for a CR-atom interaction 
to produce a positron of energy $E$. This differential cross section 
$\approx \theta(E'-E)\theta(E'-E_{\rm th}) \sigma_{ij}/E$ with $E_{\rm th}\sim 5$ 
GeV~\citep{2009A&A...501..821D}, so that if $\phi$ is described by 
a power law $\phi_0\epsilon^{-\gamma_s}$, then ${\cal Q}_s^{i,j}(E,\vec{x})
\approx 4\pi\phi_i(E,\vec{x})\,\sigma_{ij}n_j(\vec{x})/(\gamma_s-1)$. 
Moreover, since nuclei have a long range propagation scale and since the ISM 
gas does not exhibit strong spatial gradient over the kpc scale 
in the Galactic disk~\citep{2001RvMP...73.1031F}, one can further approximate 
$\phi$ and $n$ to their local values to get rough estimates of ${\cal Q}_s$. 
For instance, by considering only the proton-hydrogen interaction, and taking 
$\phi_p(E) = 1.3\,{\rm cm^{-2}s^{-1}sr^{-1}((E-m_p)/1{\rm GeV})^{-2.72}}$
\citep{2001ApJ...563..172D}, $n_H = 1 \,{\rm cm^{-3}}$ and $\sigma_{pp} = 10$ 
mb, we end up with ${\cal Q}_s (E)\approx 2\times 10^{-27}\epsilon^{-2.72}
{\rm cm^{-3}GeV^{-1}s^{-1}}$, which is actually quite close to the accurate 
calculation~\citep{2009A&A...501..821D}. Since the positron horizon is limited 
to a few kpc, the source may be modeled by an injection rate homogeneously 
distributed in an infinite thin Galactic disk of half-height $h\sim 100$ pc 
like ${\cal Q}_s(E,\vec{x})= 2h\delta(z)\,Q_{0s}\epsilon^{-\gamma_s}$. This 
allows to infer the local flux inside that disk, relevant for an observer on 
Earth:
\ben
\phi_{s,\odot}(E) = \frac{\beta c}{4\pi} \int dE_s\int d^3\vec{x}_s \,
{\cal G}_e (E,\vec{x}_\odot \leftarrow E_s,\vec{x}_s) {\cal Q}_s 
(E_s,\vec{x}_s)
= \frac{\beta c\, 2 h\,Q_{0s}}{2\pi^{3/2} b(E)} 
\int \frac{dE_s}{\lambda} \,\epsilon_s^{-\gamma_s} %\nn\\
 \approx 
\frac{c \, Q_{0s}\,h \,\sqrt{\tau} \, 
\epsilon^{-\gamma_s -\frac{1}{2}(\delta+\alpha-1)}}
{2\pi^{3/2}\sqrt{K_0}(\gamma_s-1)} \,
\;.
\label{eq:sec_approx}
\een
Here, we have taken a diffusion coefficient $K_d(E) = K_0 \epsilon^\delta$
and an energy loss rate $b(E) = (E_0/\tau)\,\epsilon^\alpha$, where $\alpha=2$ 
in the Thomson approximation. Using $\delta = 0.7$, $\alpha=2$ and 
the values given above for the other parameters, \citeeq{eq:sec_approx} gives
$\phi_{s,\odot}(E) \approx 8.5\times 10^{-3}\epsilon^{-3.57}
{\rm cm^{-2}GeV^{-1}s^{-1}sr^{-1}}$, overshooting the exact 
calculation by a factor of 2 only~\citep{2009A&A...501..821D,Del_inprep}.

\citeeq{eq:sec_approx} makes explicit the influence of the main propagation
parameters: the energy loss timescale $\tau$ and 
the diffusion coefficient normalization $K_0$ set the positron flux amplitude,
and their energy dependence slightly shapes the spectrum. Since the 
B/C ratio constrains mostly the ratio $K(E)/L$~\citep{2001ApJ...555..585M}, 
where $L$ is the vertical extent of the diffusion zone, it is not surprising 
that the {\em min} ({\em max}) model of~\citet{2004PhRvD..69f3501D}, which was 
designed to minimize (maximize) the primary antiproton flux coming from DM 
annihilation, is actually found to maximize (minimize) the secondary positron
flux. Indeed, it is associated with a small value of $L=1$ kpc (15 kpc), which 
has a corresponding small (large) value of $K_0$ to fulfill the B/C constraint.
Likewise, the logarithmic slope of the diffusion coefficient $\delta$ is 
larger in the {\em min} setup, leading to a softer spectrum than in the 
{\em max} case. These extreme configurations, all compatible with the B/C
constraints, are useful to bracket the theoretical uncertainties.

\begin{figure}[t]
  \includegraphics[width=.45\textwidth]{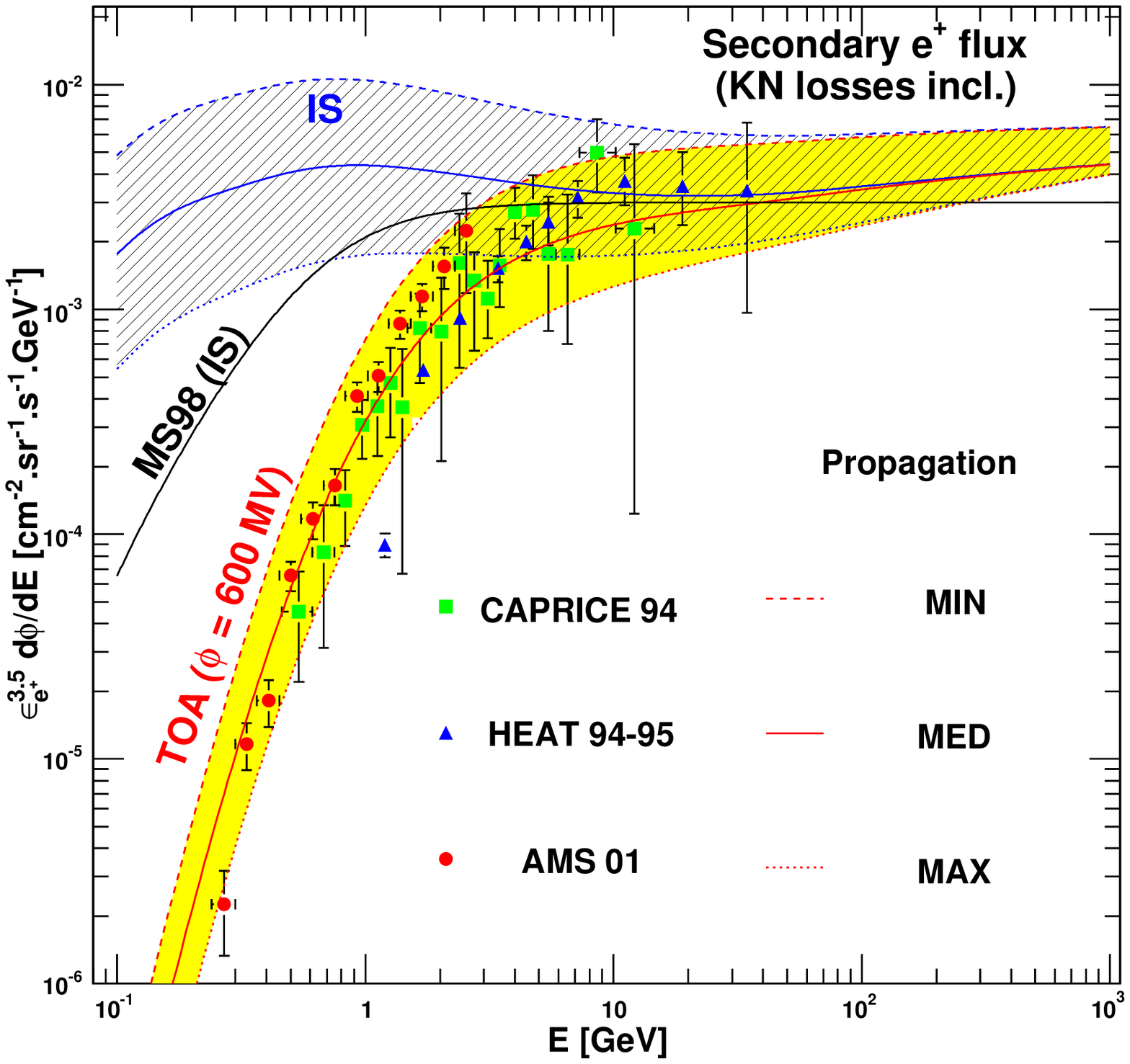}
  \includegraphics[width=.45\textwidth]{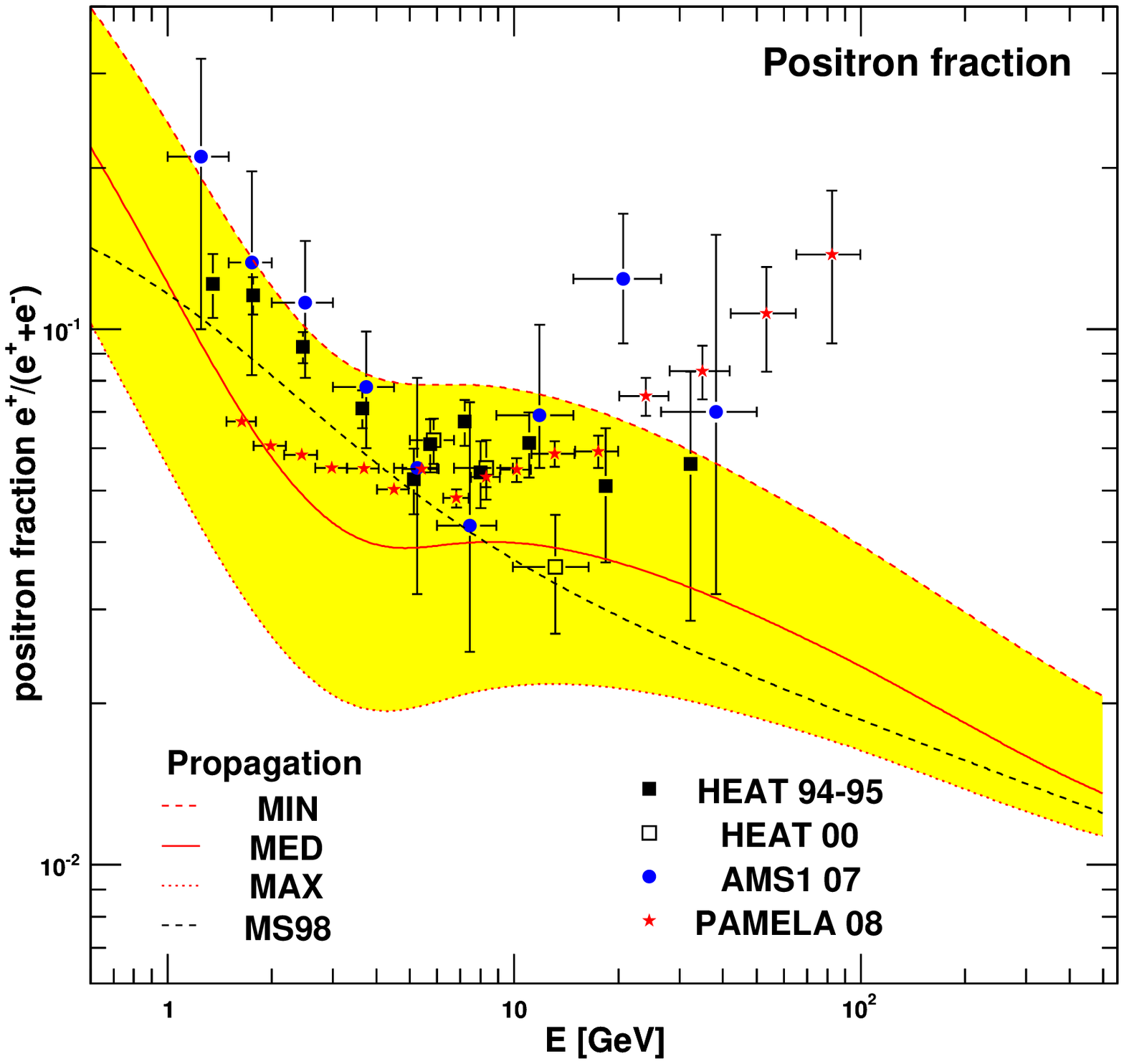}
  \label{fig:secondaries}
  \caption{Left: secondary positron flux and theoretical uncertainties. 
  Right: corresponding positron fraction. These plots are adapted from 
  \citet{Del_inprep}.}
\end{figure}

In Fig.~\ref{fig:secondaries}, we plot the latest results obtained 
by \citet{Del_inprep} for the secondary positron flux, where a full 
relativistic treatment of the energy losses, \ie~beyond the Thomson 
approximation, was used at variance with the earlier 
calculations by \citet{1998ApJ...493..694M} and \citet{2009A&A...501..821D}. 
The right panel shows the secondary flux at the Earth, where the top of 
atmosphere (TOA) signal is corrected with a Fisk potential of 600 MV to 
account for solar modulation, while the right panel is the corresponding 
positron fraction defined by $f = \phi_{e^+}/(\phi_{e^+}+\phi_{e^-})$. For the 
fraction, we fitted the electron flux on the AMS 
data~\citep{2002PhR...366..331A} below 20 GeV, and the full denominator itself 
on the Fermi data~\citep{2009PhRvL.102r1101A} above. Predictions are shown 
against the data from~\citep{2000ApJ...532..653B,2001ApJ...559..296D,2002PhR...366..331A} for the positron flux, and from~\citep{1997ApJ...482L.191B,2004PhRvL..93x1102B,2007PhLB..646..145A} for the positron fraction. For the latter, are 
also reported the recent results obtained by the PAMELA 
collaboration~\citep{2009Natur.458..607A}. We stress that though the 
theoretical uncertainties are large, of about one order of magnitude in terms 
of flux, our predictions encompass the data. However, from the spectral 
trend observed in the positron fraction, it seems unlikely that the 
excess observed by PAMELA is of secondary origin. This naturally leads 
to the question of whether or not standard astrophysical sources may provide 
enough primary positrons to explain this fraction rise, which, from 
Fig.~\ref{fig:secondaries}, should amount up to $\sim 5-10$ times the 
secondary flux around 100 GeV ($\phi_s(100\,{\rm GeV})\simeq 4\times 10^{-10}
{\rm cm^{-2}GeV^{-1}s^{-1}sr^{-1}}$).

For primary electrons and positrons, the approximations made above hold, except 
that the source term will differ. Assuming again that standard 
sources (SNRs and pulsars) are homogeneously distributed in a thin disk of 
volume $V_d = 2\pi\, h\,R^2$, the injection of CR can be written as follows:
$ %\ben
{\cal Q}_p (E,\vec{x}) \approx 2\,h\delta(z)\,
{\cal Q}_{0p}\epsilon^{-\gamma_p}e^{-E/E_c}\;,
$ %\een
where we have introduced an energy cut-off $E_c$, and where the normalization 
${\cal Q}_{0p}$, which carries the dimensions, can be fixed from energetics, 
\eg~by requiring that the total rate of injected energy is set by the supernova 
(SN) explosion rate times the energy input associated with pulsars or SNRs. For 
SNRs, we can impose that ${\cal Q}_{0p} = \Gamma_{sn} (f\,E_{sn}) / 
\{ V_d \int dE\,E \, \epsilon^{-\gamma_p}e^{-E/E_c} \}$, where $\Gamma_{sn}$ is 
the SN explosion rate in the Galaxy and $E_{sn}$ is the explosion kinetic 
energy of a single object whose a fraction of $f$ is transferred to 
electrons; for pulsars, we would use instead ${\cal E}_0$, the magnetic energy 
of which a fraction $f$ would be converted into electron-positron pairs. 
Thus, the flux of primary electrons and positrons can also be approximated with 
\citeeq{eq:sec_approx}, replacing for the normalization and keeping in mind 
that the spectral index $\gamma_p$ is different, which again turns out to be a 
fair approximation~\citep{Del_inprep}. The ratio of primaries to secondaries,
given by $R(E)=\phi_{p,\odot}(E)/\phi_{s,\odot}(E)\approx 
({\cal Q}_{0p}/{\cal Q}_{0p})\, \epsilon^{-\gamma_p+\gamma_s}$ allows to perform
a quick estimate of the pulsar contribution. With reasonable values 
$(\gamma_p,f\,{\cal E}_0,\Gamma)= (2.2,10^{47}{\rm erg},5/{\rm cy})$, we have 
$R(100\,{\rm GeV})\gtrsim 1$, which is sufficient to explain the positron 
fraction data. Anyway, even when treated more 
accurately such a modeling suffers much larger theoretical uncertainties than 
secondaries. First, the fraction of accelerated leptons and the averaged 
injected spectrum are not yet very well constrained by dynamical studies of 
sources, while important numerical efforts have been undertaken on this topic 
for a few years~\citep[\eg][]{2007ApJ...661..879E}. Second and more dramatic, 
since the explosion rate of supernov\ae~is only of a few per century in the 
whole Galaxy, the time and related spatial fluctuations become sizable locally 
and makes it difficult to justify a smooth injection rate, at least at the kpc 
scale around an observer. This is particularly relevant for the high energy 
component of the spectrum for which the typical propagation scale is short, and 
of which local sources are therefore expected to provide the main part. This is 
actually well known for decades~\citep{1970ApJ...162L.181S}, and was well 
illustrated by \eg~\citet{2004ApJ...601..340K} for electrons. An important 
consequence of these local fluctuations is that features in the local spectra 
of CR electrons and positrons are expected. Anyway, despite the very large 
uncertainties, local pulsars, which are observed in number in the solar 
vicinity and whose properties can be constrained, can inject an amount of 
positrons that is sufficient to explain a rising positron fraction. 
This was recently nicely discussed in~\citet{2009arXiv0903.1310M}. A more 
detailed study of primary electrons and positrons including local sources will 
be found in~\citet{Del_inprep}, where it is shown that all current observations
can be rather well reproduced with reasonable parameters.
%Note that {\em reproducing} does certainly not mean {\em predicting}.

To conclude this part, we stress that the background to consider
when looking for exotic signatures in the positron (or electron) spectrum
is not only made of secondaries, but also of astrophysical primaries. 
Moreover, despite the large theoretical uncertainties affecting 
current predictions, standard sources seem capable to yield the necessary 
amount of positrons that may explain the positron fraction fairly naturally, 
without any over-tuning of the parameters. Likewise, we emphasize that the 
time and spatial fluctuations of the local injection rate --- local sources --- 
can lead to a broad diversity of features in the measured 
spectrum~\citep{Del_inprep}, which makes it difficult to disentangle different 
primary components. Finally, it seems now clear that we are far from a 
{\em standard model} of Galactic CRs, especially in the lepton channel, and 
many issues remain to be addressed in the future, from the CR source 
description to a more refined propagation modeling.

\section{Dark matter and positrons}
\label{sec:dm}
Positrons were 
long thought to be good tracers of DM annihilation precisely because they were 
expected to be of secondary origin only, \ie~with a low level and 
predictable astrophysical background. As argued in the previous section, this 
statement is likely not valid anymore. Anyway, to keep the reasoning as 
general as possible, let us recall some basic conditions for a cosmic 
messenger to be a good tracer for any exotic signal: (i) the 
background is not too high with respect to the expected signal, given an 
experimental sensitivity; (ii) the background is known or predictable, and 
controled; (iii) specific spectral features in the signal make it unambiguously
distinguishable from the background. We did not yet discuss condition (i), but 
it is clear from the previous section that conditions (ii) and (iii) cannot be 
fulfilled. From this simple argument, we can hardly hope, at least with current 
data, to identify a clean DM signature in the {\em local} positron spectrum. 
Nevertheless, it is still well-grounded to ask whether DM is about to provide 
a sizable signal, should it be mixed with other components. If so, there might 
still be some hopes for isolating it with future 
experiments, provided the astrophysical background is being much better 
understood in the meantime.

We will first review the predictions that can be derived in general cases when
modeling the Galactic DM halo with a smooth distribution. Then, we will 
discuss the potential impact of DM substructures on the expected signal.
We will focus our discussion on annihilating DM, disregarding decaying 
candidates.

\subsection{The smooth approximation}
Structure formation in a $\Lambda$CDM universe involves non-linear 
processes as soon as the linear growth of perturbations triggers the 
gravitational collapse of objects. In this theoretical framework, galaxies are 
expected to have formed around redshift $z\sim 6$, and have consequently
left the linear regime for a long time, so that only numerical experiments can
provide detailed information on the DM distribution in those objects. The 
advent of high resolution numerical simulations in the study of structure 
formation has led to major breakthroughs during the last two decades, providing 
a fairly good understanding of the properties of the large scale structures 
that are observed in current surveys. There are still some important mismatches 
at the Galactic scale, but this might be due to the important impact of baryons 
which have not been included to those simulation until recently~\citep[see 
\eg][for a discussion on the small scale issues]{2009arXiv0909.2247P}. Anyway, 
disregarding the potentially large effect induced by baryons, it seems that DM 
structures at the galactic scale are predicted to have similar smooth DM 
density distributions, almost spherical and scale-independent. Such a generic 
density profile was named NFW after their authors Navarro, Frenk and 
White~\citep{1997ApJ...490..493N}, and is typified by the following function:
\ben
\rho(r) = \rho_s \frac{(r_s/r)}{(1+r/r_s)^2}\;,
\een
where $r_s$ is a scale radius beyond which the logarithmic slope goes from -1
to -3; $\rho_s$ is the scale density. It is noteworthy that more recent results 
give quite similar smooth DM components scaling like $\rho\sim(r/r_s)^{-\gamma}$ 
in the central regions, with $\gamma\approx 1$, though the lack of resolution 
prevents from making clear predictions at the very center 
\citep[see \eg][]{2008MNRAS.391.1685S,2008Natur.454..735D}. Using such a 
density shape, derived from theoretical constraints, to describe 
our Galaxy needs to account for additional observational constraints from 
stellar kinematics. Basically, although the baryon modeling comes into play 
with uncertainties, a local density of 
$\rho_\odot(R_\odot=8\,{\rm kpc})\sim 0.3\,{\rm GeV.cm^{-3}}$ associated with 
a scale radius of $r_s\lesssim 20$ kpc are fairly compatible with current data
\citep[\eg][]{2002ApJ...573..597K,2009PASJ...61..153S,2009arXiv0907.0018C}.
Now, since the injection rate of the DM annihilation products is 
$\propto\rho^2$, the precise values of the logarithmic slope may have a strong 
impact on predictions, especially for annihilation close the Galactic center 
(GC). This is particularly true for $\gamma$-ray flux 
predictions~\citep{1998APh.....9..137B}, but not really for CR positrons in the 
GeV-TeV energy range, since, as discussed above, they cannot pervade beyond 
a few kpc --- for longer range antiprotons, the signal coming from there is 
anyway diluted by diffusion. Uncertainties in the flux amplitude will be 
therefore mostly set by those on the local DM environment.

To summarize, let us write the source term associated with DM annihilation:
\ben
{\cal Q}_\chi (\vec{x},E) &=& {\cal S} \, 
\left\{\frac{\rho(r)}{\rho_\odot}\right\}^2 \, \frac{dN(E)}{dE}\;\;\;
{\rm with} \;\;\; {\cal S} \equiv \delta \frac{\sigv}{2}
\left\{\frac{\rho_\odot}{\mchi}\right\}^2 \;,
\label{eq:dm_source}
\een
where $\delta = 1/2$ (1) for Dirac (Majorana) fermionic WIMPs, and 1 for 
scalar WIMPs; $\sigv$ is the WIMP annihilation cross section, \mchi~the WIMP
mass and $dN(E)$ the number of CR positrons injected in the energy range $dE$.
The different WIMP candidates should have similar annihilation cross sections 
of $\sigv \sim 3\times 10^{-26} {\rm cm^3/s}$ if they decouple thermally from 
the primordial bath as a consequence of expansion in the early universe, such 
a value being fixed by the present cosmological DM density~\citep[the generic 
method to compute the relic abundance can be found in][]{1997PhRvD..56.1879E}. 
The local positron spectral shape will primarily depend on the annihilation 
final states. Three typical final states may basically typify the main features 
of the positron injection spectra associated with DM annihilation or decay, 
(i) quarks, say $b\bar{b}$, (ii) $W^+W^-$ and (iii) $l^+l^-$ with any charged 
lepton, say $e^+e^-$. The spectrum is getting harder and harder from (i) to 
(iii).

\begin{figure}[t]
  \includegraphics[width=.45\textwidth]{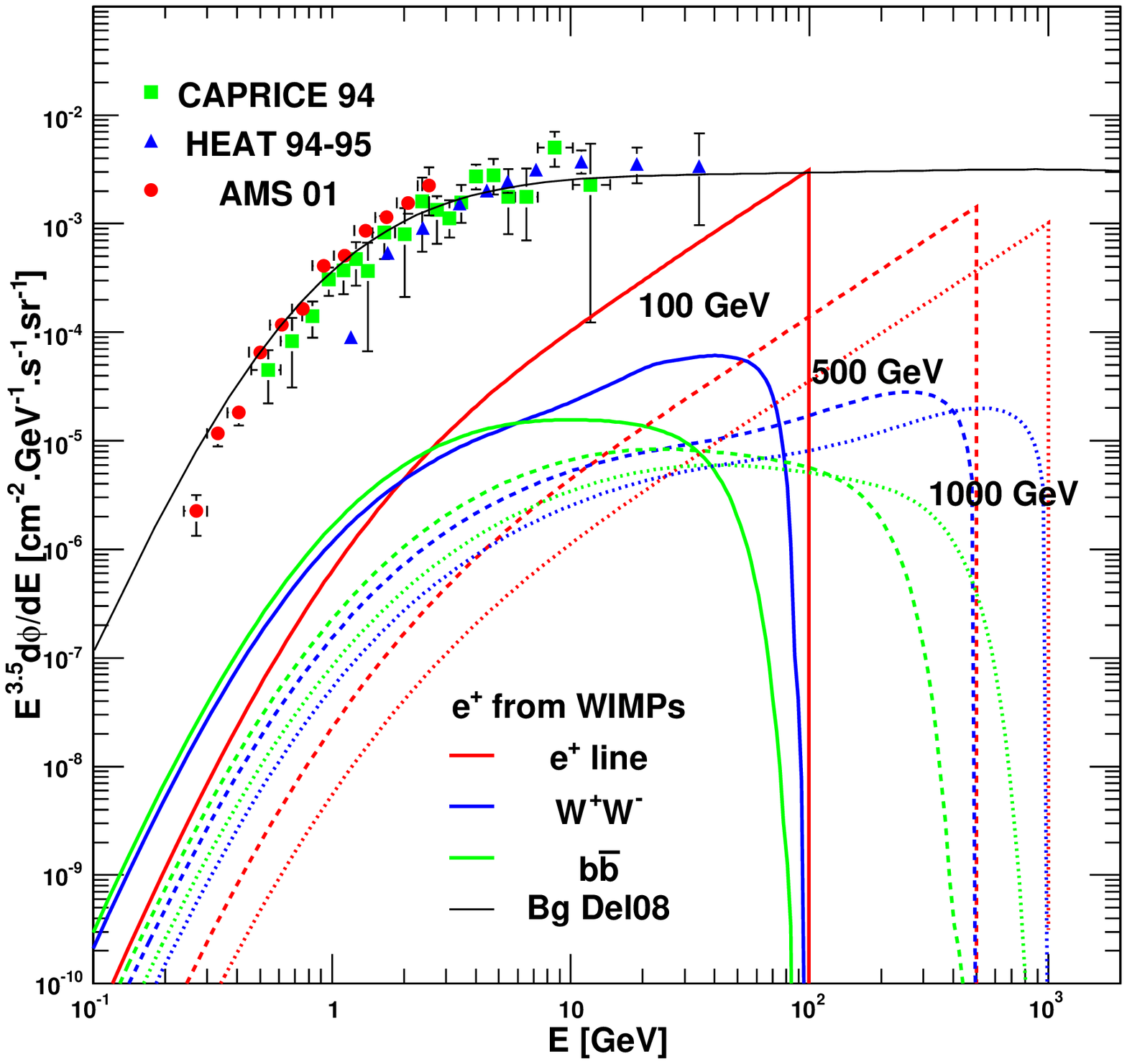}
\includegraphics[width=.45\textwidth]{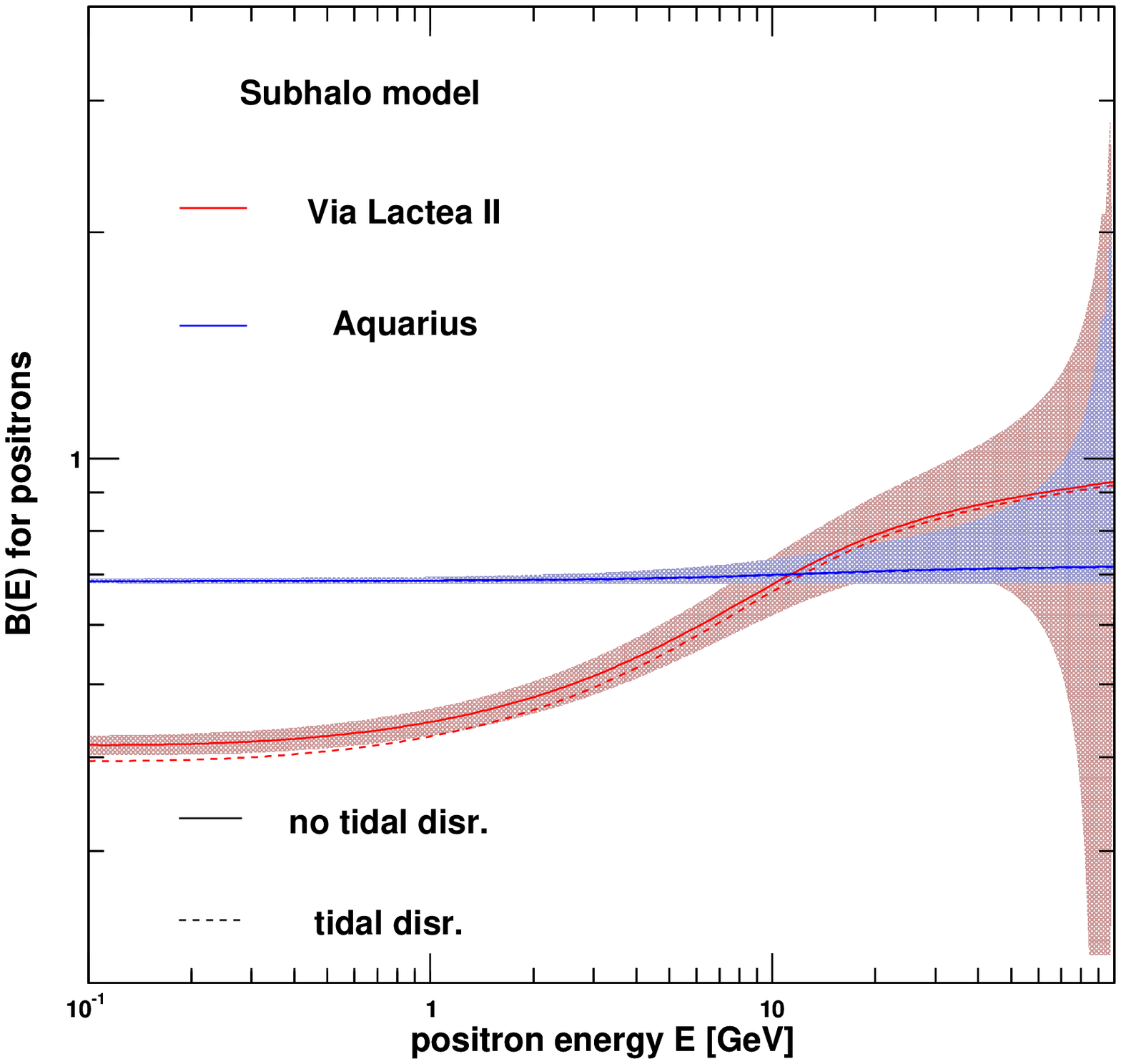}  
  \label{fig:pos_dm}
  \caption{Letf: Predictions of the primary positron fluxes for a smooth NFW 
  DM halo and 3 different annihilation final states, as compared with the 
  secondary background. Right: Boost factors obtained 
  by~\citet{2009arXiv0908.0195P} for direct annihilation in $e^+e^-$ as 
  calculated in the frames of the Via Lactea II~\citep{2008Natur.454..735D} and 
  Aquarius~\citep{2008MNRAS.391.1685S} setups.}
\end{figure}

To check whether usual WIMP candidates are about to give a sizable positron
flux, it turns out useful to derive the flux in the asymptotic limit of 
very short propagation scale, which is a rough approximation valid at high 
energy. The corresponding propagator is therefore 
${\cal G}_e(E,\vec{x}\leftarrow E_s,\vec{x}_s)
%%\underset{\lambda\rightarrow 1}{\rightarrow}
\xrightarrow{\lambda\rightarrow 0}
\delta(E_s-E) \delta^3(\vec{x}_s-\vec{x})/b(E)$ --- which only differs from 
the diffusionless limit by the term $\delta(E_s-E)$ that ensures 
$\lambda\rightarrow 0$ 
here. If we further assume a direct annihilation in $e^+e^-$, so that 
$dN(E)/dE=\delta(E-\mchi)$, then the asymptotic and {\em exact} flux limit 
reads:
\ben
\phi^{\chi}_{\odot}(E\rightarrow\mchi)
&\xrightarrow{\chi\chi\rightarrow e^+e^-}& 
\frac{c}{4\pi}\, \frac{{\cal Q}_\chi (\vec{x}_\odot,E)}{b(E)} = 
\frac{c}{4\pi}\, \frac{\cal S}{b(E)}\nn\\
&\approx& 
3.2 \times 10^{-10}{\rm cm^{-2}GeV^{-1}s^{-1}sr^{-1}} \times\nn\\
&& \frac{\sigv}{3\times 10^{-26}{\rm cm^3/s}}\, 
\frac{\tau}{10^{16}{\rm s}} \,
\left[ \frac{\rho_\odot}{0.3\,{\rm GeV/cm^3}} \right]^2\,
\left[ \frac{\mchi}{100\,{\rm GeV}} \right]^{-4} \;.
\label{eq:local_dm_flux}
\een
where we have used the Thomson approximation for the energy losses.
Note that for $\mchi=100$ GeV, this result is pretty close to the 
prediction of the secondary positron flux at 100 GeV. Since the positron 
fraction measurement at 100 GeV implies $\sim 5-10$ times more primaries than
secondaries, this means that boosting the local DM density by a factor of
$\sim 3$ is enough to feed the PAMELA data rather significantly. Nevertheless, 
this is, at least to our knowledge, the unique example for which one may 
recover the observed positron fraction at $\sim 100$ GeV without over-tuning 
the annihilation cross section. Indeed, for other annihilation final states 
and/or larger WIMP masses, one needs to boost the signal by 2 to 4 orders of 
magnitude to get enough positrons at 100 GeV to match the measurements
\citep[\eg][]{1998PhRvD..59b3511B,2008arXiv0808.3867C}. This is illustrated in 
the left panel of Fig.~\ref{fig:pos_dm}, where we compare predictions assuming 
the three annihilation final states discussed above, a smooth NFW density 
profile and the parameters used in \citeeq{eq:local_dm_flux}. In this plot, the 
secondary background is taken from~\citet{2009A&A...501..821D}, derived in the 
Thomson approximation for the energy losses.

To summarize this basic analysis, it seems that few WIMP candidates with 
thermal relic abundance may provide sufficiently positrons to feed 
the positron fraction data naturally. Indeed, only those WIMPs with masses 
around $\sim 100$ GeV and with direct annihilation in $e^+e^-$ do not need 
arbitrarily high boost factors with respect to a smooth description of the 
density profile. Needless to say that there are poor motivations for such 
models in particle physics beyond the standard model --- couplings to heavier 
leptons would lower the positron yield, however, why 100 GeV particles 
should couple only to $e^+e^-$ ? --- and that there might already exist limits 
coming from colliders.%% citation here ?
Therefore, it is fair to conclude that DM annihilation does not provide a 
natural explanation to the PAMELA data. However, it is not less fair to ask 
about the potential impact of DM substructures that are predicted in the frame 
of $\Lambda$CDM, and that could enhance the local DM density. Note that 
we have studied the impact of relaxing spherical symmetry for the smooth halo 
profile in~\citet{2008PhRvD..78j3526L}, showing that this has poor effect 
on predictions.

\subsection{Clumpiness effects}
DM substructures (called also subhalos or clumps) are predicted in the 
$\Lambda$CDM paradigm and observed in cosmological N-body simulations. The 
smallest scales that can grow and further collapse are those encompassed within 
the WIMP free streaming scale which is set by their intrinsic properties (mass,
couplings). For generic WIMPs, the minimal mass scale ranges within $10^{-11}-
10^{-3}M_\odot$ \citep{2009NJPh...11j5027B}. The mass function of these subhalos
is usually found close to the $\propto M^{-2}$ prediction of the Press-Schechter
theory of self-similar gravitational collapse 
\citep{1974ApJ...187..425P,1993MNRAS.262..627L} in cosmological N-body 
simulations~\citep[for recent results, see][]{2008Natur.454..735D,2008MNRAS.391.1685S}. Consequently, without loss of generality, the 
subhalo distribution in a Milky-Way-like object may be written 
as~\citep{2008A&A...479..427L}:
\ben
\frac{d\cl{n}(\cl{m},\vec{x})}{d\cl{m}\,dV} = 
\cl{N} \, \frac{ d{\cal P}_m(\cl{m},\vec{x}) }{d\cl{m} } \, 
\frac{d\prob_v (\vec{x} )}{dV}
\label{eq:sub}
\een
where $\cl{N}$ is the total number of subhalos, and where $d\prob_m$ and 
$\prob_V$ 
denote the mass and spatial probability distribution functions, which are 
normalized to unity over the whole galaxy extent. 
$\cl{N}$ can be constrained from N-body simulation results, at least in the 
available resolved mass range --- the most recent simulations involving 
billions of particles can 
resolve clumps down to $10^{3}-10^4\msun$ at the galaxy scale
\citep{2008Natur.454..735D,2008MNRAS.391.1685S}. When extrapolated down to a 
minimal scale of $10^{-6}\msun$ for subhalos, this number is found in the range 
$10^{14}-10^{16}$ in a Milky-Way-like galaxy~\citep{2008A&A...479..427L,2009arXiv0908.0195P}. Analytical studies on the tidal disruption of these subhalos 
in galaxies due to gravitational interactions with the disk or stars show that 
an important fraction may actually survive~\citep{2008PhRvD..77h3519B}. 
Besides, although rather spherical, the spatial distribution of subhalos turns 
out to be different from the smooth DM profile $\rho(r)$ since it exhibits a 
core radius rather than a cusp in the central parts of N-body galaxies, which 
is likely due to efficient tidal disruption, and a $r^{-2}$ form rather than 
$r^{-3}$ in the outskirts. Such a behavior is called {\em antibiased}, because 
$(d\prob_V/dV)/\rho(r)\propto r$~\citep{2004MNRAS.352..535D}, but it is 
still not clear whether it is still valid for lighter objects, far from 
being resolved in N-body simulations. Finally, the mass distribution should in 
principle depend on the location in the galaxy to account for tidal disruption.
One can include this effect by calculating the maximal subhalo mass as a 
function of the galactic radius, keeping constant the normalization of the mass 
function, which is performed in the full mass range~\citep{2009arXiv0908.0195P}.
Anyway, as mentioned above, $d\prob_m\propto \cl{m}^{-\alpha}d\cl{m}$ with 
$\alpha\lesssim 2$.

Since substructures are generically {\em predicted} in the $\Lambda$CDM 
paradigm, and are expected to be impressively numerous in galaxies if DM is 
made of WIMPs, it is important to include them for consistent calculations of 
astrophysical signals. Indeed, since the DM annihilation rate is proportional 
to the squared DM density, the presence of subhalos in the local environment 
can have strong impact on the antimatter flux (as well as on the diffuse photon 
emission). We have therefore to derive a method to add the subhalo contribution 
to the smooth one. Since we have sketched a phase-space distribution of subhalos
in \citeeq{eq:sub}, we may think about treating the flux coming from a single 
object like a stochastic variable, which actually turns out to be correct and 
powerful~\citep{2007A&A...462..827L,2008A&A...479..427L}: 
the typical range of subhalo scale radii, where most of the annihilation 
proceeds, is indeed smaller than the typical propagation scale, so that clumps 
can be safely treated as point-like sources. We still further need to specify 
the properties of a single object, \ie~its mass $m$, position $R$ 
in the Galaxy, inner density profile and the amount and spectral shape of 
positrons that it injects. It is conventional to define the subhalo extent by 
the radius $r_v$ at which the average subhalo density is 200 times the critical 
density $\rho_c$ of the universe today. Even when fixing the shape of the inner 
profile, taking \eg~the NFW model, constraining the associated scale parameters 
$r_s$ and $\rho_s$ is in principle much more complicated, since they depend on 
the formation history. Nevertheless, this history shows up an evolving 
correlation between the concentration, defined by $c_v\equiv r_v/r_s$, and the 
subhalo mass --- the less massive the more concentrated because formed 
earlier, in a denser universe. Thus, the knowledge of this concentration 
function at $z=0$ allows to specify the subhalo parameters entirely. Not only 
does this concentration function depend on the subhalo mass, but also 
on the its location in the Galaxy, since more concentrated objects resist more 
efficiently to tidal effects. It is convenient to define the {\em annihilation 
volume} of a single object:
%\ben
$
\xi (m,R) = 4\pi \int_0^{r_v} dr \,r^2\,
\left\{\frac{\cl{\rho}(m,R,r)}{\rho_\odot}\right\}^2\;.
$
%\label{eq:def_xi}
%\een
This actually defines the volume needed from a constant density of 
$\rho_\odot$ to produce the actual subhalo injection rate, and somehow measures
the ratio of its intrinsic emissivity to the local emissivity. Armed with this
definition, it is straightforward to derive the local average flux associated 
with the whole subhalo population~\citep{2007A&A...462..827L,
2008A&A...479..427L}:
\ben
\phi^{\chi}_{{\rm cl},\odot}(E) &=& \frac{\beta c}{4\pi}\,\cl{N} \,{\cal S}\, 
\langle \langle \xi(\vec{x}_s)   \rangle_m 
\wt{\cal G}(\vec{x}_\odot\leftarrow\vec{x}_s )\rangle_V
\;\;\; {\rm where} \;\;\; 
\wt{\cal G}(\vec{x}_\odot\leftarrow\vec{x}_s ) = 
\int_E^{\mchi} dE_s\,{\cal G}(\vec{x}_\odot,E\leftarrow\vec{x}_s,E_s )
\, \frac{dN(E_s)}{dE_s}\nn\\
&\xrightarrow{\lambda\rightarrow 0}& \frac{\beta c}{4\pi}\,\cl{N} \,{\cal S}\, 
\langle \xi(\vec{x}_\odot)   \rangle_m \,\frac{d\prob_V(\vec{x}_\odot)}{dV}
\, \frac{1}{b(E)}\,\frac{dN(E)}{dE}\;,
\label{eq:cl_flux}
\een
where $\langle\rangle_x$ denotes the average performed with the distribution 
$\prob_x$, and the latest line is the limit corresponding to a vanishingly 
small propagation scale $\lambda\rightarrow 0$.

In order to check whether a clumpy DM halo leads to a larger positron flux, 
it is useful to compute the ratio of both predictions. Of course, clumps
are not merely additional mass in the halo, there should be some 
consistency as well as observational constraints to obey. In fact, the census
of subhalos in N-body simulation rests on the mass resolution: since the 
galactic scale is described quite accurately, one can consider that whatever 
the discreteness of its content, an N-body galaxy will have a constant mass. 
Therefore, to model the Galaxy in a consistent manner, a certain fraction of 
mass should be removed from the smooth DM profile when adding clumps: 
$\rho(r)\rightarrow (1-f) \rho(r)$, where $f = \cl{N}\langle \cl{m} \rangle_m / 
M_{\rm MW}$. The potential problem with such a procedure is that as soon as the 
spatial distribution of subhalos differs from the smooth distribution, then the 
average local DM density is modified~\citep{2008A&A...479..427L}, which can 
lead to comparing situations with different local average DM densities. Anyway,
we can now derive the ratio of the flux prediction for the smooth case 
to that for the clumpy case, so-called {\em boost factor}:
\ben
{\cal B}_\odot(E) = (1-f)^2 + 
\frac{\phi^{\chi}_{{\rm cl},\odot}(E)}{\phi^{\chi}_{\odot}(E)}%\nn\\
\xrightarrow{\lambda\rightarrow 0} (1-f)^2 + 
\cl{N} \, \frac{d\prob_V(\vec{x}_\odot)}{dV}\, 
\langle \xi(\vec{x}_\odot)   \rangle_m \;,
\label{eq:boost}
\een
where the limit of vanishingly small propagation scale is obtained from 
\citeeq{eq:local_dm_flux} and \citeeq{eq:cl_flux}.
This expression is quite natural: the boost limit is only given by the local 
number density of objects $\cl{N} \, d\prob_V(\vec{x}_\odot)/dV$ times the 
average annihilation volume of a single object 
$\langle \xi(\vec{x}_\odot)   \rangle_m $ (which is normalized to the local 
smooth luminosity by definition, see above). Before taking 
a numerical example, we emphasize that the general expression of the boost 
factor is a function of energy. Indeed, for large propagation scales, \ie~ 
low energy, the signal coming from the cuspy smooth distribution in the 
GC will certainly dominate the total flux, while subhalos may dominate at 
short propagation scales. This is an important feature which is very often 
neglected. Notice also that the global subhalo flux is associated with a 
statistical variance that increases as the number of objects decreases in the 
relevant propagation volume: this variance does therefore increase with 
energy~\citep{2007A&A...462..827L,2008A&A...479..427L}. Correspondingly, the 
boost factor can also have a large variance provided subhalos dominate over 
the smooth contribution, otherwise it is diluted.

Let us take a very simple and rather optimistic example in which we assume a 
total number of $\cl{N}=10^{17}$ subhalos with fixed masses $10^{-6}\msun$, 
with inner NFW profiles, a fixed concentration of $c_v=100$, and distributed 
according to a cored isothermal profile of core radius $r_c=20$ kpc, extended
up to 280 kpc in the Galaxy taken with a mass of $10^{12}\msun$. We have 
therefore $d\prob_V(r_\odot=8\,{\rm kpc})/dV \approx 5\times 
10^{-7}{\rm kpc^{-3}} $, a subhalo mass fraction of $f=0.1$, so that:
\ben
{\cal B}^{-6}_\odot \approx  (1-f)^2 + \cl{N}\,\frac{d\prob_V(\odot)}{dV}\,
\frac{200\, c_v^2 (3+c_v(3+c_v))}{9(c_v+1) 
(1-\frac{(c_v+1)}{c_v}\ln(c_v+1))^2}\,
\frac{m\,\rho_c}{\rho_\odot^2} = 1.02 \;.
\een
This result is quite modest even with rather optimistic parameters
($\xi \overset{\sim}{\propto} c_v^3 m$). Should have we taken a subhalo spatial 
distribution tracking the smooth component, we would have found 
${\cal B}^{-6}_\odot\approx 4$. From this simple estimate, it seems unlikely 
that light subhalos, even if numerous, provide a strong enhancement of the 
positron signal in average. Interestingly, this naive reasoning gives a result 
which is actually quite close to accurate calculations involving more complete 
subhalo models~\citep{2008A&A...479..427L,2009arXiv0908.0195P}. For 
illustration, we show in the right panel of Fig.~\ref{fig:pos_dm} the boost 
factors and associated statistical variances obtained from the subhalo settings 
derived in \citet{2008Natur.454..735D} and \citet{2008MNRAS.391.1685S}, where 
WIMPs of 100 GeV annihilating in $e^+e^-$ are assumed. It is noteworthy that in 
this plot, boost factors are shown to be in fact less than 1!  This is 
a consequence of the decrease in the local density due to the addition of 
subhalos, constrained so to keep the Galaxy mass constant. The 
statistical variance shown as colored bands expresses the fact that few 
nearby objects can actually dominate the whole subhalo contribution. This 
would induce peculiar features in the positron spectrum, hardly distinguishable
from those predicted for nearby standard astrophysical sources
\citep[\eg][]{2009arXiv0907.5093R}.

\section{Conclusion}
\label{sec:concl}

We have argued that the rising local positron fraction observed by the PAMELA 
satellite is unlikely of secondary origin. Then, we have discussed the 
potential yield from astrophysical sources of primary positrons, 
emphasizing local pulsars as our best candidates, and stressing that those 
conclusions were already sketched twenty years ago~\citep{1989ApJ...342..807B}.
Although the overall electron and positron data can be very well explained with 
standard astrophysical processes, we have finally stressed that we are still 
far from a {\em standard model} of cosmic rays, since the current theoretical 
uncertainties on propagation, source modeling as well as ISM modeling make it 
rather unfair to claim for clean {\em predictions}; at the moment, we can only 
provide rough estimates still {\em tuned} --- though reasonably --- to reproduce
the data. In light of this discussion, it seems to us that the rising positron 
fraction, as well as the so-called {\em electron excess} sometimes {\em seen} 
in the Fermi data, are no longer theoretical issues, since standard and not 
contrived explanations are available. Remains wide open the question of 
identifying and modeling accurately the local sources of cosmic ray leptons in 
order to sustain this solution on more detailed grounds. These are rather good 
news for this research domain, which, besides the need for important theoretical
efforts, can benefit a copious amount of experimental data. Of course, this 
implies a multimessenger and multiwavelength analysis, which are mandatory for 
consistency purposes.

Regarding the DM hypothesis, we have shown that usual WIMP candidates are not
expected to contribute significantly to the local positron flux, even when 
treated in a self-consistent framework including subhalos. The only possibility 
without over-tuning the annihilation cross section allowed for thermal relics 
is to consider direct production of $e^+e^-$ and a mass scale of $\sim$ 100 GeV,
which is not motivated in particle physics theories beyond the standard model. 
Likewise, we have also stressed that should DM yield a sizable positron signal, 
it would be difficult to disentangle it from standard astrophysical sources. 
This illustrates the fact that the basic conditions that would make positrons 
good DM tracers are not fulfilled: not only is the background larger than the 
signal, but, more important, it is not yet under control.

Anyway, we underline that though unlikely contributing to the local positron 
flux, WIMPs remain excellent DM candidates. The crucial issue of their 
detection is still challenging, since their expected properties have made them 
continuously escape from observation despite the advent of important 
experimental devices, especially in high energy astrophysics. It seems 
important to develop more complex strategies based on multi-messenger, 
multi-wavelength and multi-scale approaches, in which large efforts should be 
made to quantify and minimize the associated theoretical uncertainties. Other 
detection methods are also very important, among which the LHC results are 
particularly expected.

{\bf Acknowledgments:} It is a pleasure to thank T. Delahaye and R. Lineros
for sharing a great deal of work in this topic.

%\begin{theacknowledgments}

%\end{theacknowledgments}

\bibliographystyle{aipproc}   % if natbib is available

\bibliography{lavalle_bib}

\IfFileExists{\jobname.bbl}{}
 {\typeout{}
  \typeout{******************************************}
  \typeout{** Please run "bibtex \jobname" to optain}
  \typeout{** the bibliography and then re-run LaTeX}
  \typeout{** twice to fix the references!}
  \typeout{******************************************}
  \typeout{}
 }

\end{document}